\documentclass{article}

\usepackage{arxiv}

\usepackage[utf8]{inputenc} 
\usepackage[T1]{fontenc}    
\usepackage{hyperref}       
\hypersetup{colorlinks=true,citecolor=cyan,urlcolor=blue}

\usepackage{url}            
\usepackage{booktabs}       
\usepackage{amsfonts}       
\usepackage{nicefrac}       
\usepackage{microtype}      
\usepackage[euler]{textgreek}
\usepackage{graphicx}
\graphicspath{ {./images/} }

\title{Acousto-optically driven single-shot ultrafast optical imaging}

\author{
 Mohamed Touil, Saïd Idlahcen, Rezki Becheker, Denis Lebrun, Claude Rozé, Ammar Hideur, and Thomas Godin \\ \\
CORIA UMR 6614 - CNRS - Universit\'{e} de Rouen Normandie - INSA Rouen, 76800 Saint Etienne du Rouvray, France\\
  \url{thomas.godin@coria.fr} \\
}

\begin{document}

\maketitle

\begin{abstract}
Driven by many applications in a wide span of scientific fields, a myriad of advanced ultrafast imaging techniques have emerged in the last decade, featuring record-high imaging speeds above a trillion-frame-per-second with long sequence depths. Although bringing remarkable insights in various ultrafast phenomena, their application out of a laboratory environment is however limited in most cases, either by the cost, complexity of operation or by an heavy data processing. We then report a flexible single-shot imaging technique combining sequentially-timed all-optical mapping photography (STAMP) with acousto-optics programmable dispersive filtering. The full control over the acquisition parameters is enabled via the spectro-temporal tailoring of the imaging pulses in an electrically-driven spectral phase and amplitude shaper in which the pulse shaping in both the temporal and spectral domains is controlled through the interaction of the light field with an acoustic wave. Here, contrary to most single-shot techniques, the frame rate, exposure time and frame intensities can be independently adjusted in a wide range of pulse durations and chirp values, making the system remarkably versatile and user-friendly. The imaging speed of the system as well as its flexibility are validated by visualizing ultrashort events on both the picosecond and nanosecond time scales. With the perspective of real-world applications and to achieve the highest technical simplicity, we eventually demonstrate its lensless operation based on digital in-line holography. The virtues and limitations as well as the potential improvements of this on-demand ultrafast imaging method are critically discussed.  
\end{abstract}

\section{Introduction}
Capturing the transient dynamics of ultrashort events focuses the attention of the scientific community for decades in biomedical science, chemistry and physics \cite{Hu06,Diebold13,Lei18,Yang18}, and ultrafast imaging is now routinely used for research and industrial applications. The need for high temporal and spatial resolutions has then remarkably fueled unprecedented advances in ultrafast optical imaging, and sub-picosecond and sub-nanometer resolutions are now accessible to various technologies. Transient phenomena are traditionally captured using pump-probe methods, which are however intrinsically limited to highly repeatable experimental conditions \cite{Mikami16,Liang18,Liang20}. Some techniques, such as time-stretch imaging \cite{Goda09}, leverage their high throughput capabilities for recording high-speed processes on long - microsecond - time scales with frame rates in the order of a hundred million of frames-per-second (fps) \cite{Wu17,Hanzard18}. Nevertheless, such frame rates are not sufficient to resolve many ultrafast events and single-shot techniques with frame intervals in the picosecond range - or less - are required for imaging e.g. laser-induced phenomena \cite{Liang17SCIADV,Yao21} or light propagation itself \cite{Liang18LSA,Ehn17}. Single-shot methods can be separated between passive and active detection and further between direct imaging and reconstruction imaging \cite{Qi2020,Liang18}, each family having its strengths and limitations. Passive detection refers to the use of custom-designed pulse trains to probe ultrashort events while active detection relies on specific ultrafast detection schemes. On the one hand, the best performances to date in terms of imaging speed and sequence depth are undoubtedly obtained with passive detection and image reconstruction using advanced algorithms such as compressed sensing. Among those computational imaging methods, compressed ultrafast photography (CUP) \cite{Gao14} and its noteworthy upgradings and variants \cite{Liang17SCIADV,Liang18LSA,Lu19,Liang20SPCUP,Yang20,Kim20} then manage to reach up to 70 trillion Hz frame rates \cite{Wang20} and theoretically up to more than 180 Tfps \cite{Qi2021}, few hundreds of femtosecond frame intervals and potentially sequences with hundreds of frames, in a single camera exposure. Albeit being the spearhead of ultrafast imaging techniques and particularly relevant in a laboratory environment, these techniques could suffer from their complex reconstruction scheme hindering real-time operation and from their low flexibility. On the other hand, active imaging techniques using spatially-, temporally-, and/or spectrally-tailored ultrafast pulse trains have enabled ultra-high frame rates with simpler experimental setups and without the need of complex computational algorithms\cite{Ehn17,Kakue12,Li14,Yue17} but, in most cases, to the detriment of sequence depth and without the possibility to record the spectrum (e.g. fluorescence) emitted from the dynamic scene. Among them, sequentially timed all-optical mapping photography (STAMP) \cite{Nakagawa14}) and more specifically its spectrally-filtered variant (SF-STAMP) \cite{Suzuki15,Suzuki17} offer trillion-fps rates with real-time operation but suffer, as most of the above-mentioned techniques, from the absence of adaptability of its main parameters - namely, frame rate and exposure time - that cannot be independently adjusted.\\
To answer this need for an independent and user-friendly "on-demand" control of these parameters for potential real-world applications, we introduce here an agile technique combining spectrally-filtered STAMP with compact acousto-optics-based electronically-controllable phase and amplitude filters. The latter, known as acousto-optic programmable dispersive filters (AOPDF) have originally been designed to compensate for the group delay dispersion (GDD) in ultrafast laser systems \cite{Tournois97}, carrier-envelope phase (CEP) stabilization \cite{Canova09,Forget09} or pulse shaping in various applications \cite{Maksimenka10,Seiler15,delaPaz19}. AOPDF have also been used as ultrafast delay lines in terahertz spectroscopy \cite{Urbanek16} and pump-probe spectroscopy and imaging \cite{Audier17}. Their ability to fully tailor the pulse phase and amplitude in a simple manner then enabled the synchronization and chirp control of fs pulses, e.g. in stimulated Raman spectroscopy \cite{Alshaykh17,Audier20}. We then propose here to leverage the AOPDF pulse shaping capabilities in a SF-STAMP ultrafast imaging scheme, and to demonstrate the independent control of frame rate, exposure time and frame intensities as well as the ultrafast imaging of ultrashort events on different time scales. The potential refinements and limitations of the method in a real-world environment are eventually discussed.\\

\section{Results}
\begin{figure}[ht]
\centering
\includegraphics[width=0.8\columnwidth]{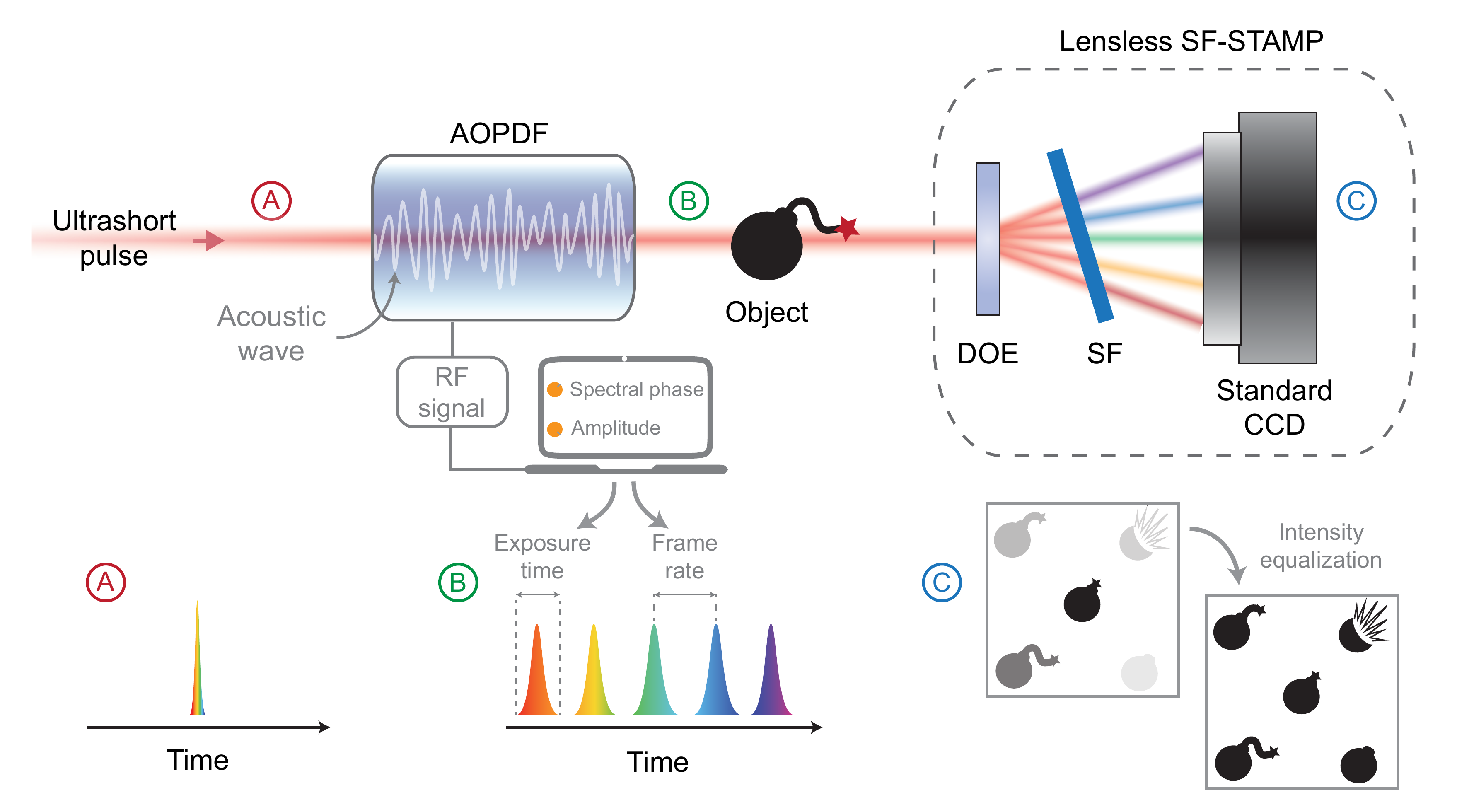}
\caption{\textbf{Principle of operation.} The acousto-optics programmable dispersive filter (AOPDF) tailors the pulse shape in both the spectral and temporal domains via its interaction with an electrically-driven acoustic wave, then enabling the full and independent control over the exposure time and frame rate in the subsequent SF-STAMP detection scheme, comprising a diffractive optical element (DOE),a tilted spectral filter (SF) and a standard camera. }
\label{fig:setup} 
\end{figure}

\paragraph{AOPDF-based sequentially timed all-optical mapping photography: principle and parameters control\\}
~~\\
In spectrally-filtered sequentially timed all-optical mapping photography experiments, broadband chirped pulses are used to illuminate the object under study, each frequency then capturing snapshots of the dynamic scene at different instants \cite{Nakagawa14}. The different wavelengths are subsequently spatially separated using the combination of a diffractive optical element (DOE) and a tilted spectral filter (SF) and the generated sub-pulses, corresponding to the consecutive frames, are recorded using a standard camera \cite{Suzuki15}. As the snapshots are obtained through the spatial separation of specifically selected spectral bands from the main spectrum of the laser pulse, the exposure time of the technique is then defined by the spectral width of the extracted sub-pulses, and by the initial chirp of the broadband input pulse. The time between frames is also determined by the initial chirp along with the spectral gap between consecutive sub-pulses. For standard SF-STAMP, the DOE divides the input pulse into several replicas propagating in different directions while, for each replica, the angle of incidence on the spectral filter fixes the transmitted central wavelength. The limited bandwidth of the input ultrashort pulse intrinsically limits the freedom in the azimuthal and radial angles of the spectral filter, and thereby hinders the acquisition of equally-spaced non-overlapping (in the spectral domain) sub-pulses. Prior to the illumination of the object, pulses are then temporally stretched and spectrally shaped in order to achieve the desired frame rates and exposure times. The pulse stretching step usually consists in propagating in glass rods or dispersive elements such as prisms while the spectral broadening and/or shaping is obtained via nonlinear propagation (SPM) in optical fibers or using more complex setups based on 4-f lines with spatial light modulators \cite{Nakagawa14,Suzuki17}. Such pulse shaping stages, in addition to being bulky, exhibit a very low flexibility as the frame rate and exposure time are fixed and cannot be independently controlled by the end-user. Here, we then circumvent these steps using a single acousto-optics programmable dispersive filter (AOPDF) for both temporal and spectral shaping and then for the full control of the illumination conditions, as shown in Fig. \ref{fig:setup}. It consists of a centimeter-long birefringent crystal where the optical pulse interacts with a radio-frequency-driven acoustic wave, enabling the simultaneous and independent tailoring of its spectral phase and amplitude. The wavelength-dependent group delay and amplitude are then controlled within a compact, turn-key device, allowing to precisely tune both the exposure time and frame rate in real-time, but also to guarantee that the diffracted replicas exhibit comparable intensities and thereby to maximise the camera dynamics (lower panel in Fig. \ref{fig:setup}).

We highlight the flexibility of our concept with the example of amplitude and phase shaping shown in Fig. \ref{fig:operation}. The input laser pulse spectrum is compared with the optimal transfer functions of the SF for each of the sub-pulses (five replicas in our case) in Fig. \ref{fig:operation}a, corresponding to an optimized set of azimuthal and radial angles. Although the five transfer functions fall within the spectral range of the input pulse spectrum, there is considerable overlap - spectral and temporal - between some of them and only three snapshots can therefore be acquired. The other major drawback is their low overlap with the input spectrum and then the huge contrast between the central image at 802 nm and the other ones, resulting on a very poor intensity dynamics, as shown in \ref{fig:operation}b. Considering a linearly chirped input pulse, this also results in a non-constant time between frames.

In order to overcome these limitations, we then performed an AOPDF-based amplitude and phase control. In the AOPDF crystal, the laser pulse interacts with a control field in the acoustic domain through a frequency mixing process where the energy is quasi-conserved. When the phase-matching conditions between the acoustic and optical waves are reached, the optical field is diffracted from the ordinary to the extraordinary axis of the crystal at the location of interaction and the intensity of the diffracted light will be proportional to that of the corresponding phase-matched acoustic wave. By properly tailoring the control field, the output spectrum can thereby be precisely shaped together with the location in the crystal where each component is diffracted. Therefore, a predetermined delay is created between the selected spectral bands, which can be fully controlled until a maximum value defined by the length of the crystal and the refractive index difference between its axes. Fig. \ref{fig:operation}c shows the amplitude mask designed to generate the desired control acoustic field and the subsequent shaped output spectrum featuring quasi-equalized spectral bands. This allows to exploit the full dynamic range of the sensor (see Fig. \ref{fig:operation}d) while preventing any spectral or temporal overlap between the five sub-pulses. 

In order to increase the diffraction efficiency of the frequency mixing process, the designed acoustic wave is set to cover most of the crystal length when interacting with the unchirped input pulse. This allows to efficiently obtain a linearly chirped output pulse with an equivalent accumulated dispersion D = 0.269 ps/nm (blue line in Fig. \ref{fig:operation}e). Based on this accumulated dispersion and on the spectral width of each sub-pulse, the exposure time (i.e. the pulse width of the sub-pulses) is estimated to an average value of 723 fs. At this stage, the time between frames corresponds to the separation between the consecutive sub-pulses. In order to even this time between frames, the position at which each of the five sub-pulses is diffracted within the crystal is adjusted via an additional delay (positive or negative) resulting in the temporal shift of the selected sub-pulses (red curve in Fig. \ref{fig:operation}e and Fig. \ref{fig:operation}f). We then obtain a single-shot ultrafast imaging system with an average exposure time of 723 fs and time between frames of 2 ps between the two first images and 1.33 ps between the next ones. We then overcome one of the main limitation of the STAMP technique as the time between the five frames can be independently equalized while the exposure time is left unchanged. This remarkable agility is obtained only through a single user-friendly electronically-controlled pulse shaping system and without resorting to any additional spectral broadening and temporal stretching stages.

\paragraph{Ultrafast imaging of an optical Kerr gate\\}
~~\\
The temporal capabilities of ultrafast systems are usually validated by imaging femtosecond- or picosecond-scale light-induced phenomena such as the propagation of a pump pulse through a Kerr medium. In order to demonstrate that our flexible technique is on par with state-of-the-art active imaging methods, we then used our system to capture the opening and closing of a CS$_2$-based optical Kerr gate (OKG). The principle of an OKG relies on placing a transparent medium with a high nonlinear refractive index (liquid CS$_2$ here) between two crossed polarizers along the optical axis of an image beam. An intense ultrashort pump pulse is then used to induce a local birefringence which results in a change of the polarization state of the image beam from linear into elliptical, therefore allowing the partial transmission of the image beam through the second polarizer. The gate time response depends on the pump pulse duration in addition to the relaxation time of the nonlinear medium. OKGs are thereby simple and efficient time gates and are routinely used e.g. for capturing ballistic photons (and reject multi-scattered photons) in ultrafast imaging experiments \cite{Idlahcen09}.

We then used our AOPDF-based system to capture the whole OKG process using the setup depicted in Fig. \ref{fig:fullsetup}a,d. A linearly polarized fs source is used for both the imaging via the pulse shaping stage and the CS$_2$ pumping via an optical delay line (ODL). On the pump arm, the ODL allows to synchronize the pump and imaging pulses and a cylindrical lens is used in order to increase the intensity of the pump within the CS$_2$. For each optical delay (step of 400 $\mu$m), the imaging system then captures five single-shot images of the OKG, as shown in Fig. \ref{fig:fullsetup}e where the images are displayed in a matrix in which the raws represent the single-shot images while the columns represent the group optical delay. For a delay of 6.6 ps, the five frames capture the whole pump-induced process, from the opening to the closing of the time gate. An horizontal spatial displacement of the intensity can also be noticed, which corresponds to the propagation of light within the CS$_2$. In order to further confirm the ability of our system to provide accurate measurements, the propagation of the pump and imaging pulses within the CS$_2$ cell have been simulated using a validated standard model \cite{Idlahcen09}. The latter allows to predict the transmitted light from the second polarizer as a function of the pump and image pulses parameters by modeling the CS$_2$ response using a single relaxation process (see “Methods” section). The simulated snapshots for each spectral component, in full agreement with the experimental measurements, are shown in Fig. \ref{fig:fullsetup}f.

\paragraph{Extending the frame interval to the nanosecond\\}
~~\\
As a lot of transient or non-repeatable phenomena (e.g. in light-matter interactions) also occur on the nanosecond time scale, and to illustrate the flexibility of our AOPDF-based imaging technique, we demonstrate its extension to nanosecond frame intervals. The extension of the sub-ps initial temporal resolution of SF-STAMP to the ns range has already been demonstrated e.g. using free-space angular-chirp-enhanced delay (FACED) \cite{Wu17,Suzuki20}, but to the detriment of simplicity as the frame rate and exposure time are intrinsically related in such techniques. Here we propose to extend the frame interval through the increase of the group delay between the five sub-pulses while keeping their respective pulse width close to the Fourier limit, thereby minimizing the ratio between exposure time and time between frames.

Here, the accessible time scale is intrinsically fixed by the length of the acousto-optic crystal (a few cm) and thereby limited to a few picoseconds. Therefore, we added the extension shown in Fig. \ref{fig:fullsetup}a to the imaging system for generating a controlled group delay with an extended and adjustable temporal window. In order to keep the system compact and easy to align, we used a modified single grating pulse compressor comprising a diffraction grating, a roof mirror retroreflector for horizontal displacement and five moving mirrors for each of the five sub-pulses.

As the input laser beam is linearly polarized, a half-wave plate is used prior to a polarizing beam splitter (PBS)  to inject the initial pulse either directly in the AOPDF for ps-scale imaging or towards the grating-based extension to reach longer time scales. The orientation of the PBS is chosen so that the polarization of the exiting beam toward the AOPDF is appropriate for the frequency mixing process. A quarter wave plate is also used to ensure that the back-propagating beam is efficiently coupled into the AOPDF after the phase modulation of the input beam. The modified single grating pulse compressor induces a negative group velocity dispersion (GVD) as an ordinary single grating compressor, and the distance between the grating and the roof mirror determines the induced GVD in addition to defining the horizontal spacing between the sub-pulses after the second reflection from the grating. Using five spatially-separated moving mirrors allows to generate a group delay between each spectral band eventually resulting in a predetermined time difference between the five imaging sub-pulses at the output of the AOPDF. In this process, the AOPDF plays the role of refining the spectrum by eliminating any existing spectral overlap between the sub-pulses. It also allows to equalizes their amplitude as described in the previous section. In addition, the positive GVD induced by the AOPDF partially compensates for the negative GVD experienced in the compressor, then minimizing the average pulse duration of the sub-pulses.  As a result, this simple modified setup allows to extend the time scale of the imaging system to the nanosecond regime while keeping the exposure time inferior to one picosecond. In order to validate its nanosecond-scale imaging capabilities, we then applied the modified technique to the tracking of laser-induced shock waves (SW). The latter are generated by focusing high-energy 532 nm nanosecond pulses on solid samples using a lens with f = 12 mm, as shown in Fig. \ref{fig:fullsetup}b, and their evolution is captured using the standard SF-STAMP system. The SW images captured for metals and glass samples are shown in Fig. \ref{fig:fullsetup}c and have been obtained without any post-processing. From the consecutive snapshots, SW velocities can then be easily inferred (e.g. a velocity of 16,26 km/s for the aluminium case). This demonstrates the remarkable versatility of the system, whether operating in the picosecond or nanosecond regime.  

\paragraph{Lensless imaging based on digital in-line holography\\}
~~\\
In our system, as in most of single-shot imaging techniques, the dynamic scene is captured through a combination of optical elements such as lenses, e.g. to access a given field of view. Such lenses however require a careful alignment from the end-user and ultrafast imaging techniques could gain in simplicity if this step was removed. With this in mind, the use of digital holography is then a natural solution. In a different scope, digital holography has previously been directly used as a space-division light-in-flight recording technique \cite{Kakue12}. This technique exhibited Tfps frame rates, but was somehow limited in terms of spatial resolution and required a reference pulse. Here we then propose to combine the AOPDF-based SF-STAMP technique with digital in-line holography (DIH) as it does not require any additional optical component nor a reference arm. All the lenses were thereby removed from the setup and holograms, such as the one shown in Fig. \ref{fig:holo}a, were recorded on the CCD sensor. Note that the hologram in Fig. \ref{fig:holo}a was simply normalized by the background intensity distribution previously recorded without the plasma excitation beam. From this hologram, the complex amplitude can be reconstructed in any plane along the optical axis using an adapted model (see "Methods") as shown in Fig. \ref{fig:holo}b, where the real and imaginary parts of the amplitude are displayed. In this experiment, ns pulses from a frequency-doubled Nd-YAG laser were focused (using a lens with f = 20 cm) to generate laser-induced air breakdown, which manifests as the specific patterns shown in Fig. \ref{fig:holo}b. This phenomenon can be explained as follows. When the pulses are focused within a small volume, it initiates laser-accelerated electrons through a multiphoton ionization process, leading to successive cascaded ionizations of the medium by inelastic collisions. This eventually results in the formation of a plasma expanding asymmetrically toward the pump laser \cite{POKRZYWKA2012}. This type of pattern has been observed using classical burst imaging techniques in the ps regime \cite{Suzuki15}, which only allows to visualize one state of this process (corresponding to the second snapshot in Fig. \ref{fig:holo}b). Here, the birth and growth of this pattern can clearly be seen in the consecutive snapshots in the reconstructed images obtained through the lensless AOPDF-based technique. From these real and imaginary part images, the phase can also easily be calculated using $\phi(x,y)=arctan(Im(x,y)/Re(x,y))$, as shown in \ref{fig:holo}b. Remarkably, calculating the phase is an efficient way to precisely determine the axial coordinate $z$ at which the object of interest (here, the highest electron density region) is located using the following procedure. From the complex amplitude images, the phase images are calculated at every axial coordinate and stacked horizontally in order to plot an orthogonal view of the phase map such as the one shown in Fig. \ref{fig:holo}c. The phase profile along the optical axis (red solid line) exhibits a typical shape that has been observed previously \cite{Shaw:06}, which is a clear signature of the object position and then gives a simple and efficient tool for an accurate focusing on the best reconstructed image.

\section{Discussion}
\paragraph{Limitations and potential enhancements\\}
~~\\
The SF-STAMP technique on which relies our system has proven efficient for capturing ultrafast events in a simple and user-friendly manner. However, its sequence depth (i.e. the number of images acquired in a single-shot) is intrinsically limited by two factors. On the one hand, 
as it exploits bandwidth-limited laser pulses together with a spectrum sampling process, having a larger number of images would necessitate the filtering and separation of very narrow adjacent spectral bands, which is feasible but could require a more complex setup. In that case, a specific attention should be paid to the potential temporal overlap between the extracted spectral bands. On the other hand, the number of images is also limited by the spatial separation of the sub-pulses (through the DOE and tilted spectral filter) prior to the acquisition on the CCD sensor. As a single CCD sensor is used, the larger the number of images, the lower the number of pixels per image and, consequently, a compromise between the sequence depth and the spatial resolution has to be found. With a view to provide a compact and simple system, the above-mentioned drawbacks could be easily suppressed using several technical solutions leveraging the AOPDF spectro-temporal tailoring capabilities, as it allows the generation of adjacent non-overlapping narrow (sub-nm) spectral bands. It would then be possible to use a broadband fiber laser as the illumination source to increase the number of shaped sub-pulses and an hyperspectral camera to replace the DOE, spectral filter and CCD sensor at once. The number of images would therefore be determined by the number of spectral bands and be independent of the spatial resolution. This would also allow to eliminate the path difference between the sub-pulses due to the DOE that results in a shift along the z-axis when reconstructing the images using DIH. 

In this work, the AOPDF-based system exhibits a fixed maximum temporal window of 8.5 ps regardless of the number of images. It is worth noting that longer crystals than the one used in this feasibility demonstration could be used. This would enable larger temporal windows (through larger GDD values) in the picosecond regime. In the nanosecond regime, the temporal window is enlarged using a modified single grating compressor which is however bulky and necessitates a careful alignment. Still in the aim of building a compact alignment-free system, the generation of the additional delay between the pulses could be performed using custom-made fiber Bragg gratings (FBG) adapted to the desired imaging temporal window. 
Regarding holography, it is clear that a rigorous phase map over the whole field of view cannot be estimated - as in off-axis holography - without a perfectly known reference \cite{picart2015}. It is also worth noting that the phase distortions in an in-line configuration could be reduced by a phase conjugation technique as shown in Ref. \cite{mazumdar2020}. However, these two potential refinements would entail a much more complex setup, which is out the scope of this study aiming for a user-friendly technique. Here, as far as a small volume is studied around the considered object of interest, in-line holography is completely satisfactory provided that the local wavefront variations remain acceptable for extracting an exploitable phase signature, as shown in Fig. \ref{fig:holo}.

\paragraph{Summary\\}
~~\\
By leveraging the AOPDF's pulse shaping capabilities within a burst imaging scheme, our single-shot technique stands out of the current state-of-the-art as its main features are easily controllable and adjustable by the end-user through a simple electrical signal. The outcomes of this feasibility demonstration are manifold and could help to bring such ultrafast techniques out of a laboratory environment. \textit{(i)} The AOPDF acts as an "all-in-one" compact and flexible device that can tailor the phase and/or the amplitude of the input signal on a wide range of pulse duration and chirp values. Unlike most of the existing techniques, the frame rate and exposure time can then be independently adjusted without resorting to complex or bulky spectro-temporal shaping stages. \textit{(ii)} The technique can be easily switched between different timescales as demonstrated here by either recording the dynamics of an optical Kerr gate in the ps regime or the evolution of laser-induced shock waves in the ns regime. \textit{(iii)} The setup can be simplified even more by removing the lenses and efficiently using digital in-line holography to reconstruct the object on a wide depth of field. In addition, we confirmed here that the axial phase profile of the reconstructed images is a striking signature of the object position on the longitudinal axis. Extra refinements could now be added at different levels on the technique (shaping stage, sequence depth, detector) in order to provide a fully-integrated alignment-free system for real-world applications requiring Tfps imaging without using any advanced computational imaging method.

\section{Methods}
\begin{figure}[ht]
\centering
\includegraphics[width=1\columnwidth]{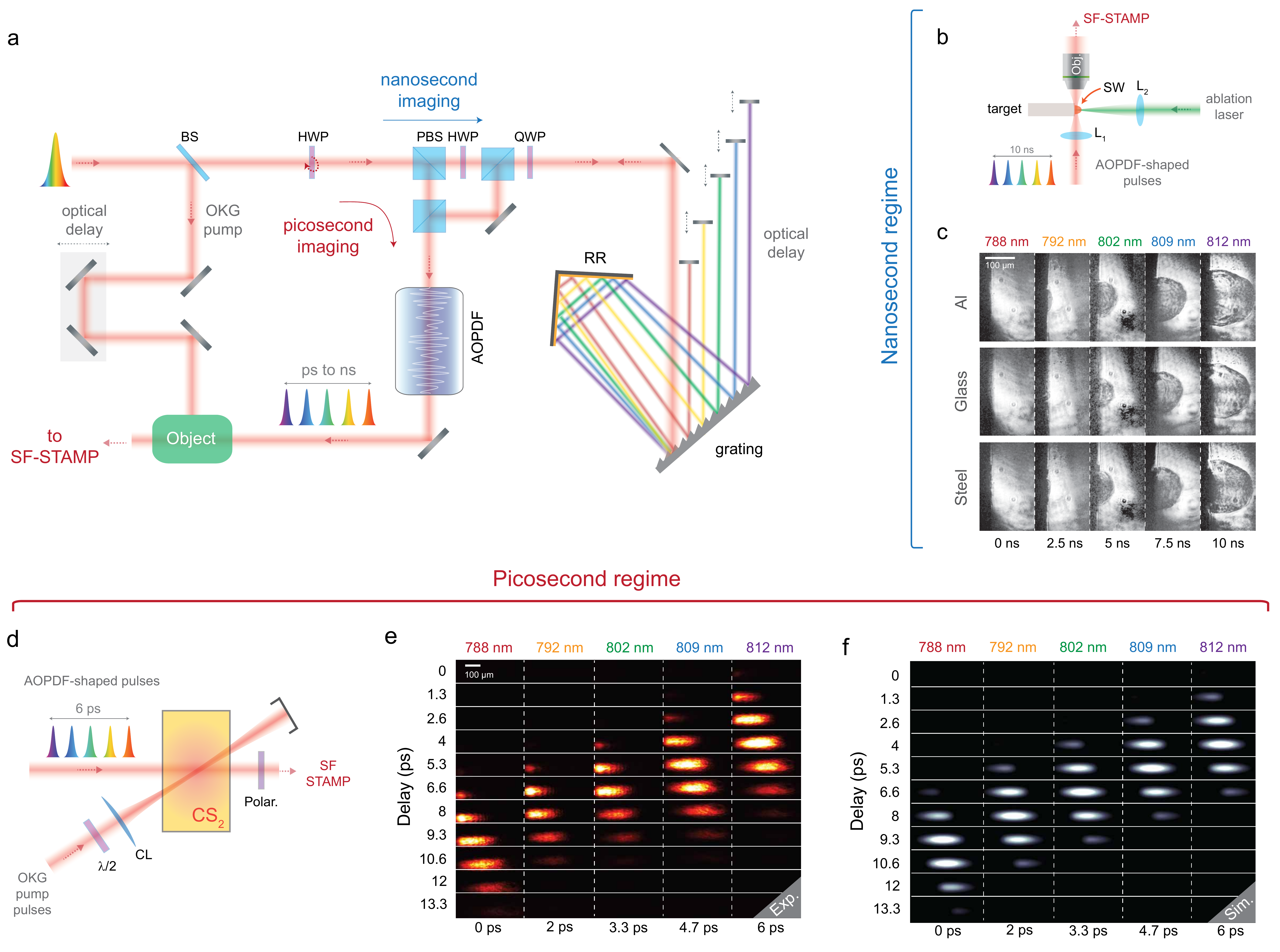}
\caption{\textbf{AOPDF-based ultrafast imaging on different timescales}. \textbf{a} Experimental setup. An half-wave plate allows to either operate the system in the picosecond or nanosecond regime. The latter is obtained using a grating-based stretcher with an additional adjustable delay between the sub-pulses prior to the AOPDF spectro-temporal shaping. \textbf{b} Experimental configuration used to generate and capture laser-induced shock waves (SW) on the nanosecond regime. SW are generated by focusing intense visible pulses onto solid targets. \textbf{c} Images of SW obtained for aluminium, glass and stainless steel without any post-processing. \textbf{d} Experimental configuration used for the picosecond-scale imaging of an optical Kerr gate (OKG) in a CS$_2$ cell. An optical delay line is used to synchronize the pump pulses with the imaging pulses in the CS$_2$ cell and to precisely capture the Kerr gate dynamics. \textbf{e} Vertically-stacked images of the transmitted pulses while varying the optical delay. The opening and closing of the OKG are fully acquired for a delay of 6.6 ps. \textbf{f} Numerical simulation of the OKG imaging process. }
\label{fig:fullsetup} 
\end{figure}

\paragraph{AOPDF-based SF-STAMP\\}
~~\\
In the experimental setup shown in Fig. \ref{fig:fullsetup}a, we used a chirped pulse amplifier (Coherent, Inc.) to generate linearly-polarized 100 fs pulses centered at 800 nm with a 12 nm bandwidth (FWHM) and a repetition rate of 1 kHz. The beam has a diameter of 3 mm and a maximum average power of 30 mW. A half-wave plate is rotated to operate the imaging system either in the picosecond or the nanosecond regime. A polarizing beam splitter (PBS) is set so that the beam polarization on the AOPDF is orthogonal to the diffraction plane of the acousto-optic crystal. In the picosecond regime, the half-wave plate is rotated so that the laser beam goes directly toward the AOPDF in order to create five pulses with equal intensities, equal time delay between them and with specifically selected central wavelengths and spectral widths. In the nanosecond regime, the half-wave plate is rotated so that the laser beam exits the PBS toward the nanosecond stage. A quater-wave plate ensures that the returning beam exits the PBS toward the AOPDF. The nanosecond stage consists of a modified single grating compressor where the laser beam is diffracted by a 600 lines/mm grating set at an angle of 45$^{\circ}$ with respect to the incoming beam's optical axis. The first-order diffracted beam is sent back to the grating by a roof mirror reflector set at 300 mm from the grating, thus providing a GVD of -1.28 ps/nm. The five selected spectral bands are then reflected back to the grating using five mirrors with a 375 mm distance between them. The first mirror reflects the spectral band up to 790 nm while the rest is left to propagate toward the remaining mirrors. In ascending order, these mirrors reflect the spectral bands up to 800 nm, 807 nm, 811 nm and the last mirror reflects the remaining part of the spectrum. A group delay of 2.5 ns between the sub-pulses is then obtained prior to injection in the AOPDF. Here, the spectral widths and central wavelengths of the five sub-pulses can be finely tuned in order to avoid any spectral overlap on the interference filter and their respective intensities can also be equalized. Finally, the GVD induced by the modified single grating compressor can be partially compensated in order to compress the five pulses closer to their Fourier limits.

Before the AOPDF, the beam diameter was reduced to 1.6 mm with a divergence inferior to 0.04$^{\circ}$ in order to maximize both the resolution and efficiency of the pulse shaping system. The AOPDF (Dazzler, Fastlite) consists of a 25 mm long TeO$_{2}$ birefringent acousto-optic crystal  for spectro-temporal pulse-shaping. Here, it allows a chirping factor up to 85 (chirped pulses up to 8.5 ps) and the shaping process is performed on a spectral range of 25 nm providing a maximum of 110 channels corresponding to a maximum resolution of 0.23 nm. In the AOPDF crystal, the longitudinal interaction between the acoustic wave and the optical field polarized along the ordinary axis can be viewed as a three wave mixing process. This interaction results in the diffraction of optical wavelengths to the extraordinary axis depending on the phase matching conditions. These conditions are given by:
\begin{equation}
\textbf{k}_{diff}= \textbf{k}_{opt} + \textbf{k}_{ac}
\label{entire_phasematch}
\end{equation}
where $\textbf{k}_{diff}$, $\textbf{k}_{opt}$ and $\textbf{k}_{ac}$ are the diffracted, optical and acoustic wavevectors, respectively. Depending on the angles of incidence of the acoustic wave and the optical field, the phase matching conditions create an almost bijective relationship between the optical and acoustic frequencies (thick crystal). This relationship is nearly linear and mainly depends on the crystal properties in addition to the phase matching conditions and is given by:
\begin{equation}
\omega_{opt}/ \omega_{ac}\sim\alpha
\label{alp}
\end{equation}
where $\alpha$ is the optical to acoustic frequencies ratio. Here, $\alpha = 2.3\times10^{-7}$. This enables pulse shaping by simply using an electronically-generated acoustic wave. The shaping process in the AOPDF can be viewed as a convolution of the complex electric field of the input optical wave with the acoustic signal. In the frequency domain, this convolution can be expressed as:
\begin{equation}
E_{diff}(\omega_{opt})= E_{in}(\omega_{opt}).S(\omega_{ac})= E_{in}(\omega_{opt}).S(\alpha . \omega_{opt}) 
\end{equation}
On the one hand, the amplitude of the diffracted electric field $E_{diff}$ at a given angular frequency $\omega_{opt}$ is proportional to the input electric field $E_{in}$ at the same angular frequency with a proportion determined by the amplitude of the acoustic wave with the frequency $\alpha$  $\omega_{opt}$ as dictated by the phase matching conditions. On the other hand, the spectral phase of the diffracted optical pulse is simply the addition of the initial phase of the input optical pulse with the induced phase due to the acousto-optic interaction in the crystal.
\begin{equation}
\phi_{diff}=\phi_{opt} + \phi_{AOPDF} 
\end{equation}
$\phi_{AOPDF}$ is a mere exploitation of the crystal birefringence to create a group delay between the wavelengths of the optical pulse depending on the position of the crystal at which they have been diffracted from the ordinary to the extraordinary axis. The group delay is given by:
\begin{equation}
\tau(\omega)= (n_{go} (\omega)/c)\times z(\omega)+(n_{ge}(\omega)/c) \times (L-z(\omega))
\end{equation}
With $n_{go}$ and $n_{ge}$ being the ordinary and extraordinary group refractive indices, respectively. c is the speed of light, $z(\omega)$ is the distance propagated by the optical frequency $\omega$ along the ordinary axis and L the total length of the crystal (L = 25 mm). By adjusting $z(\omega)$, the group delay of every optical frequency $\omega$ and the phase of the diffracted optical pulse can thereby be controlled. It is worth noting that the shaped pulse eventually exits the acousto-optic crystal at an angle of 1.4$^{\circ}$ with respect to the input pulse.

In the shock waves generation experiment, images were acquired on a small area using a microscope system comprising a f = 50 mm condenser lens and an objective lens ($\times$20, NA = 0.35, Nachet). The beam was subsequently enlarged by a factor two using another telescope. The DOE and SF are eventually comprised within a $4f$ imaging system with a magnification of one.

In the SF-STAMP detection, a diffractive optical element (DOE, Holoeye DE224) allows to generate 2$\times$2 beams at a diffraction angle of 5.1$^{\circ}$ in addition to the non-diffracted beam. In order to select a different spectral band on each of the diffracted beams, a tilted band-pass spectral filter (BPF, IRIDIAN, ZX000167) is set after the DOE. It exhibits a transmittance superior to 90 $\%$ at 800 nm and a spectral bandwidth $\Delta\lambda_{SF}$ of 2.2 nm. The BPF is adequately tilted in such a way to allow a selective transmission of each incident spectral component at a specific location on its surface. Here, the azimuthal and radial angles are found to optimal at $\theta$ $\sim$ $21^{\circ}$ and $\varphi$ $\sim$ $3^{\circ}$ knowing that the central wavelength of the SF is angle-dependent and, for an angle $\beta$ with respect to the normal of the SF plane, is given by \cite{Suzuki15,Suzuki17}:
\begin{equation}
\lambda(\beta)=\lambda_0\left(1-\frac{\beta^2}{2n_{eff}^2}\right)
\label{bandpass}
\end{equation}
From this, the corresponding wavelength-dependant transmitted intensity can be expressed as:
\begin{equation}
I(\lambda)=exp\left(\frac{-4ln(2)(\lambda-\lambda(\beta))^2}{\Delta\lambda_{SF}^2}\right)
\label{intensitywavelength}
\end{equation}
Finally, the five sub-pulses are eventually captured using a standard CCD camera (JAI, RM-4200 CL) with 2048$\times$2048 pixels and 15.15$\times$15.15 mm sensor dimensions. 

\begin{figure}[ht]
\centering
\includegraphics[width=1\columnwidth]{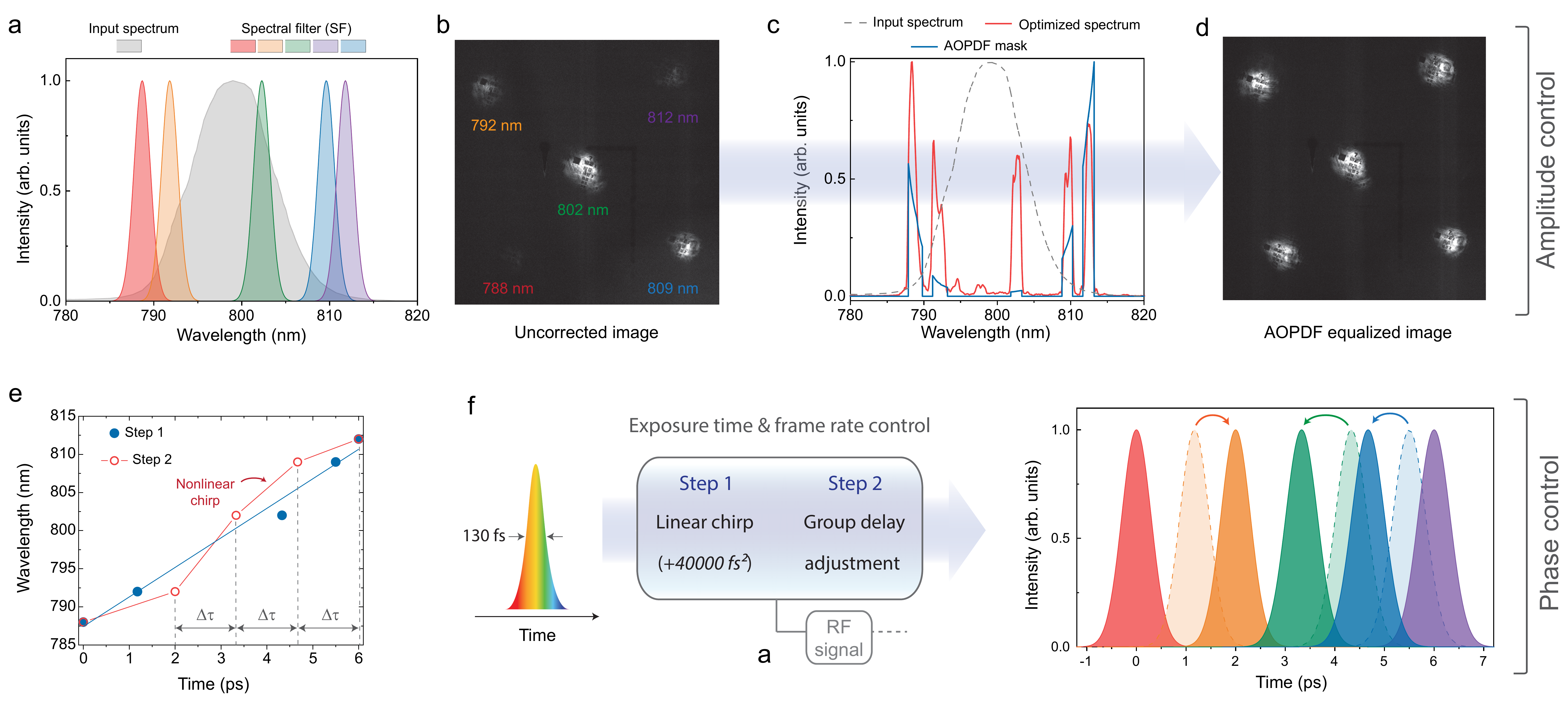}
\caption{\textbf{AOPDF-based amplitude and phase control.} \textbf{a} Input laser spectrum compared with the normalized transmission bands of the STAMP spectral filter (SF). \textbf{b} The convolution of the input spectrum with the SF leads to an inhomogeneous sub-pulse intensity distribution on the CCD camera and is associated with poor intensity dynamics. \textbf{c} Amplitude mask applied on the AOPDF (blue) and resulting optimized spectrum (red) compared with the input spectrum (dotted gray). \textbf{d} Resulting image with equalized intensities and optimized dynamics. These are raw images without any post-processing. \textbf{e},\textbf{f} Spectro-temporal distributions of the illuminating pulses. Exposure time and frame rate can be independently controlled by adding a linear chirp and adjusting the group delay of each spectral component, respectively. The arrows correspond to the group delay effect on the sub-pulses.}
\label{fig:operation} 
\end{figure}

\paragraph{Calculation of the system's performances\\}
~~\\
In the picosecond time scale, the temporal resolution of our system can be designed and controlled through the sole exploitation of the AOPDF's crystal birefringence. We begin by introducing an accumulated dispersion D = 0.269 ps/nm on a spectral range of $\Delta\lambda_{window}$ = 25 nm to obtain an observation window of $\Delta{T}$ using the following relation:
\begin{equation}
\Delta{T}= D.{\Delta{\lambda_{window}}}
\label{Time_window_obs}
\end{equation}
The temporal characterization is performed through the imaging of an optical Kerr gate (OKG) in a pump-probe experiment. The image beam size was reduced  (using a pinhole) to limit the spatial interaction in the CS2 cell with the pump beam. The transmittance of the OKG in time is a function of the delay between the pump and image pulses. Therefore, since the proposed acousto-optic filtering results in the generation of 5 independent temporally separated pulses, the transmittance of the OKG for each of the five pulses is identical and is delayed in time with a delay corresponding to the time between the pulses (time between frames). This time can be easily obtained using cross-correlation. Using the time separation between the pulses and their corresponding central wavelength, a first order linear regression is used to estimate the accumulated dispersion parameter D as shown in Fig. \ref{fig:operation}e. Once the accumulated dispersion D is determined, along with the optical spectrum we are able to obtain the pulse width (exposure time, ET) of each of the five pulses using the following expression \cite{Nakagawa14}:
\begin{equation}
ET(i)= \sqrt{\left(\frac{2{\lambda}^{2}_{0i} {ln\,2}}{\pi c \Delta \lambda_{i}} \right)^{2}+(D.\Delta \lambda_{i})^{2}},
\label{exposure_time}
\end{equation}
where \textit{i} is the pulse number, $\Delta \lambda_{i}$ is the spectral bandwidth and $\lambda_{0i}$ is the central wavelength. The time between frames is further adjusted by adding a positive or negative group delay to the 2$^{nd}$, 3$^{rd}$ and 4$^{th}$ pulse to equalize the time separation between our five pulses. This obviously does not affect their pulse duration. Using the same pump probe experiment, we can define the time separation between the five pulses and after few iterations we obtained a quasi-constant time between frames as shown in Fig. \ref{fig:operation}e,f. In the nanosecond time scale, the accumulated dispersion is the summation of the contribution of the modified single grating compressor and the AOPDF. Its value is estimated to be $D = 1.01 $ ps/nm. Therefore, the average pulse duration in this case is ${ET_{average} \sim 2}$ ps. As for the time between frames, it was precisely adjusted to 2.5 ns.

\paragraph{OKG imaging\\}
~~\\
In our experiment, the femtosecond laser pulse was linearly polarized and split by a 90/10 beam splitter prior to the pulse shaping stage as shown in Fig. \ref{fig:fullsetup}a. 90\% of the power was used as a pump while the rest was directly injected into the AOPDF. At the output of the AOPDF, we obtained five linearly polarized pulses with equal intensity and temporal and spectral separations as discussed in the previous section. A CS$_2$ cell and a single polarizer were eventually set between the AOPDF and the lensless SF-STAMP as a Kerr time-gate. The pump pulses were synchronized with the image pulses via an optical delay line and a cylindrical lens (f = 75 mm) was used in order to focus the pump beam on a single axis (vertical) inside the 10 $\times$ 10 mm CS$_2$ cell at an angle of 42$^{\circ}$ with respect to the image beam.

\paragraph{Theoretical model for the optical Kerr gate\\}
~~\\
In the simulation of the CS$_2$-based OKG, we considered the change of the refractive index in time and space due to the pump intensity as follows \cite{Idlahcen09}:
\begin{equation}
\Delta{n}(\textbf{r},t)=n_{2}\frac{\tau_{0}+\tau_{r}}{\tau_{0}^2} \int_{-\infty}^{t}I_{p}(\textbf{r},\tau) exp\left[-\frac{t-\tau}{\tau_0}\right]\left[1-exp\left[-\frac{t-\tau}{\tau_r}\right]\right]d\tau  
\label{delta_n} 
\end{equation}
with $t \geq 0$, where $\tau_0$ is the rise time ($\tau_{0}=0.14$ ps), $\tau_r$ the relaxation time ($\tau_{r}=1.0$ ps), $n_2$ the nonlinear refractive index of CS$_2$ ($n_{2}=3.0 \times 10^{-19} m^{2}/W$) \cite{Idlahcen09} and $I_p (\textbf{r} ,\tau)$ the intensity of the pump in time and space. The pump beam is given an elliptical spatial profile and a Gaussian temporal shape corresponding to an intensity distribution given by:
\begin{equation}
I_{p}(\textbf{r},t)=I_{pmax} exp \left[  -\frac{x^{2}}{x_{p}^2}-\frac{y^{2}}{y_{p}^2}-\left(\frac{t}{\tau_p}\right)^2\right] 
\label{pumppulse}
\end{equation}
with $x_{2}$ = 3.8 mm (corresponding to a FWHM = 5.3 mm), $y_2$ = 1.4 mm, $\tau_{p}=60 fs$ (FWHM = 100 fs). The energy per pulse is E = 0.4 mJ and $I_{pmax}$ (in W/m$^2$) is the maximum intensity of the pump. The image is composed of 5 pulses delayed in time, which all have the same Gaussian shape, in space and time. Nevertheless, as there is no overlap between the pulses on the detector, only one pulse propagation is simulated, and is delayed by a time $\tau_{d}$ to mimic the other pulses. The probe pulse is thus written as:
\begin{equation}
I_{im}(\textbf{r},t)=I_{immax} exp\left[-\frac{x^{2}+y^{2}}{r_{im}^2}-\left(\frac{t-t_d}{\tau_{im}}\right)^2\right]
\label{imagepulse}
\end{equation}
where $r_{im}$ = 1.0 mm (FWHM = 1.6 mm), $\tau_{im}=0.41$ ps (FWHM = 0.723 ps), $\tau_{d}=0.0,2.0,3.3,4.7,6.0$ ps and $I_{immax}$ is the maximum intensity of the image pulses. The CS$_{2}$ cell is rectangular and much larger than the overlap zone of pump and probe pulses. The length of the cell in the image beam propagation direction \textit{z} is 10 mm. The angle between the two beams is $\Theta= 42^{\circ}$ in air, corresponding to $26^{\circ}$ inside the cell according to the Snell-Descartes laws. The pump beam induces a birefringence at an angle of $45^{\circ}$ relatively to the \textit{x} and \textit{y} axes. The phase change of the image pulse along a propagation distance $\delta {z}$, where image and pump pulses overlap, is given by:
\begin{equation}
\delta\varphi=\frac{2\pi\delta{z}}{\lambda} \Delta{n}
\label{phasechange}
\end{equation}
For each time interval, the overlap between the image and pump pulses is computed and the phase change $\delta{\varphi}$ is deduced. The overall transmitted intensity is finally given by Eq. \ref{kerrtransmitance}, where $\varphi$ is the accumulation of all $\delta\varphi$.
\begin{equation}
I=sin^2\frac{\varphi}{2}
\label{kerrtransmitance}
\end{equation}

\paragraph{Digital in-line holography (DIH)\\}
~~\\
The recording process of digital in-line holograms can be simply expressed by applying a propagation model along the z-axis (i.e. the optical axis of the imaging system). Let $A_{im} (\textbf{r} ,t)$ be the complex amplitude function due to the probe pulse $I_{im} (\textbf{r} ,t)$ and considered as the object to be recorded. The complex amplitude distribution in the CCD sensor plane located at a distance $z_e$ from the object is obtained from the Huygens-Fresnel integral. According to Ref. \cite{Onural93}, and by omitting the constant multiplicative phase term $exp\left(\frac{2i\pi z_e}{\lambda}\right)$, this integral can usefully be expressed as the following 2-D convolution :

\begin{equation}
A_{holo}(\textbf{r},t,z_e)=\left[1-A_{im}(\textbf{r},t)\right]**h(\textbf{r},z_e)
\label{hologram_amplitude}
\end{equation}

where $h(\textbf{r},z_e)$ is the Fresnel kernel :
\begin{equation}
h(\textbf{r},z_e)=\frac{1}{i\lambda{z_e}}exp\left[\frac{i\pi(x^2+y^2)}{\lambda{z_e}}\right].
\label{fresnelkernel}
\end{equation}

The intensity distribution $\left(=A_{holo}.\overline{A_{holo}}\right)$ recorded by the sensor can also be expressed by using this formalism. Knowing that $1**h=1$, it gives :

\begin{equation}
I_{holo}(\textbf{r},t,z_e)=1-A_{im}(\textbf{r},t)**\left[h(\textbf{r},z_e)+\overline{h(\textbf{r},z_e)}\right]
\label{hologram_intensity}
\end{equation}

where the top bar $\bar{.}$ stands for the complex conjugate.

Note here that for simplification, the squared modulus term $|A_{im}(\textbf{r},t)**h(\textbf{r},z_e)|^2 $ has been omitted. This approximation is valid provided that the far-field conditions are satisfied  \cite{Buraga2000}.

As for the recording step, the reconstructed image at a given distance $z_r$ from the CCD sensor can also be calculated by a convolution operation :  

\begin{equation}
R(\textbf{r},t,z_r)=I_{holo}(\textbf{r},t,z_e)**h(\textbf{r},z_r)
\label{hologram_reconstruction}
\end{equation}

It is easy to see, by introducing Eq. \ref{hologram_intensity} in \ref{hologram_reconstruction}, that when the right reconstruction distance is reached (i.e. $z_r=z_e=z_{opt}$), we have :

\begin{equation}
R(\textbf{r},t)=1-A_{im}(\textbf{r},t)-A_{im}(\textbf{r},t)**h(\textbf{r},2z_{opt}).
\label{reconstructed_image_amplitude}
\end{equation}

\begin{figure}[ht]
\centering
\includegraphics[width=1\columnwidth]{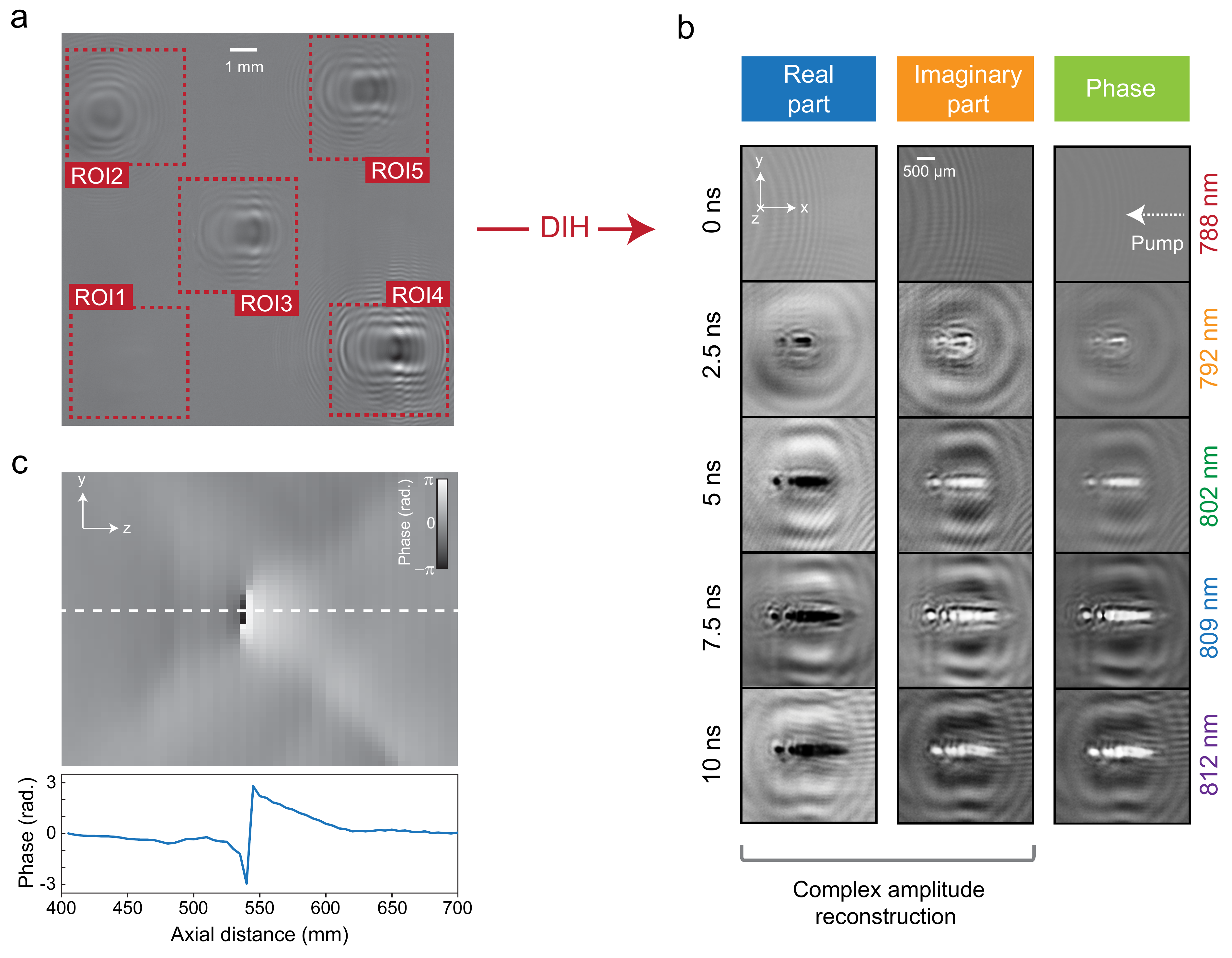}
\caption{\textbf{Image reconstruction using digital in-line holography (DIH).} The concept of lensless AOPDF-based SF-STAMP based on DIH is demonstrated by imaging laser-induced air breakdown on the ns scale. \textbf{a} Hologram normalized by background subtraction. \textbf{b} Complex amplitude and phase reconstruction using DIH (based on the hologram of \textbf{a}). The image is reconstructed at an axial distance z$_r$ = 550 mm and the asymmetrical plasma expansion is clearly seen. \textbf{c} Phase map along the longitudinal axis obtained by reconstructing the complex amplitude and calculating the phase at different longitudinal coordinates. The selected zone of interest is the leftmost spot in the air breakdown pattern in \textbf{b}. The phase variation along the optical axis (blue solid line) is a striking signature of the actual location of the intensity and phase object.}
\label{fig:holo} 
\end{figure}

The reconstructed complex image amplitude $1-A_{im}(\textbf{r},t)$ is surrounded by the waves coming from the so-called twin image $-A_{im}(\textbf{r},t)**h(\textbf{r},2z_{opt})$.
In the present case, the far-field conditions are achieved and we assume that the contrast of the twin image fringes does not disturb the reconstructed sample volume. Examples of hologram recording (see \ref{fig:holo}a) and reconstruction (see \ref{fig:holo}b) are presented. The phase values of $R(\textbf{r},t)$ - here mapped in the $(y,z)$ plane - are shown on Fig. \ref{fig:holo}c. It is noticeable that this orthogonal phase map representation is quite similar to the phase signature presented in \cite{Shaw:06}.  The phase discontinuity observed here allows efficient detection of the best focusing plane in the reconstruction volume. In the example shown on Fig. \ref{fig:holo}c, the object axial coordinate has been estimated to $z_{opt}=539 mm$.   
  
\paragraph{Acknowledgements\\}
~~\\
This work was supported by the French Agence Nationale de la Recherche and Labex EMC3, the European Union with the European Regional Development Fund, and the Regional Council of Normandie (TOFU and IFROST projects).

\paragraph{Author contributions\\}
~~\\
MT, SI, RB, AH and TG designed and built the system, performed the experiment and analyzed the data. CR provided the optical Kerr gate model. DL performed the image reconstruction using digital in-line holography. MT, RB, AH and TG wrote the manuscript. All authors contributed to the discussion of the results. 

\paragraph{Conflict of interest\\}
~~\\
The authors declare that they have no conflict of interest.

\bibliographystyle{unsrt}  
\bibliography{references}  

\begin{thebibliography}{10}

\bibitem{Hu06}
S.~X. Hu and L.~A. Collins.
\newblock Attosecond pump probe: Exploring ultrafast electron motion inside an
  atom.
\newblock {\em Phys. Rev. Lett.}, 96:073004, Feb 2006.

\bibitem{Diebold13}
Eric~D. Diebold, Brandon~W. Buckley, Daniel~R. Gossett, and Bahram Jalali.
\newblock Digitally synthesized beat frequency multiplexing for sub-millisecond
  fluorescence microscopy.
\newblock {\em Nature Photon.}, 7:806--810, 2013.

\bibitem{Lei18}
Cheng Lei, Hirofumi Kobayashi, Yi~Wu, Ming Li, Akihiro Isozaki, Atsushi
  Yasumoto, Hideharu Mikami, Takuro Ito, Nao Nitta, Takeaki Sugimura, Makoto
  Yamada, Yutaka Yatomi, Dino~Di Carlo, Yasuyuki Ozeki, and Keisuke Goda.
\newblock High-throughput imaging flow cytometry by optofluidic time-stretch
  microscopy.
\newblock {\em Nature Protocols}, 13:1603--1631, 2018.

\bibitem{Yang18}
Jie Yang, Xiaolei Zhu, Thomas J.~A. Wolf, Zheng Li, J.~Pedro~F. Nunes, Ryan
  Coffee, James~P. Cryan, Markus G{\"u}hr, Kareem Hegazy, Tony~F. Heinz, Keith
  Jobe, Renkai Li, Xiaozhe Shen, Theodore Veccione, Stephen Weathersby, Kyle~J.
  Wilkin, Charles Yoneda, Qiang Zheng, Todd~J. Martinez, Martin Centurion, and
  Xijie Wang.
\newblock Imaging cf3i conical intersection and photodissociation dynamics with
  ultrafast electron diffraction.
\newblock {\em Science}, 361(6397):64--67, 2018.

\bibitem{Mikami16}
Hideharu Mikami, Liang Gao, and Keisuke Goda.
\newblock Ultrafast optical imaging technology: principles and applications of
  emerging methods.
\newblock {\em Nanophotonics}, 5(4):497--509, 2016.

\bibitem{Liang18}
Jinyang Liang and Lihong~V. Wang.
\newblock Single-shot ultrafast optical imaging.
\newblock {\em Optica}, 5(9):1113--1127, Sep 2018.

\bibitem{Liang20}
Jinyang Liang.
\newblock Punching holes in light: recent progress in single-shot
  coded-aperture optical imaging.
\newblock {\em Reports on Progress in Physics}, 83(11):116101, oct 2020.

\bibitem{Goda09}
K.~Goda, K.~K. Tsia, and B.~Jalali.
\newblock Serial time-encoded amplified imaging for real-time observation of
  fast dynamic phenomena.
\newblock {\em Nature}, 458:1145--1149, 2009.

\bibitem{Wu17}
Jiang-Lai Wu, Yi-Qing Xu, Jing-Jiang Xu, Xiao-Ming Wei, Antony~CS Chan,
  Anson~HL Tang, Andy~KS Lau, Bob~MF Chung, Ho~Cheung Shum, Edmund~Y Lam,
  Kenneth~KY Wong, and Kevin~K Tsia.
\newblock Ultrafast laser-scanning time-stretch imaging at visible wavelengths.
\newblock {\em Light Sci. Appl}, 6:e16196, 2017.

\bibitem{Hanzard18}
Pierre-Henry Hanzard, Thomas Godin, Saïd Idlahcen, Claude Rozé, and Ammar
  Hideur.
\newblock Real-time tracking of single shockwaves via amplified time-stretch
  imaging.
\newblock {\em Appl. Phys. Lett.}, 112(16):161106, 2018.

\bibitem{Liang17SCIADV}
Jinyang Liang, Cheng Ma, Liren Zhu, Yujia Chen, Liang Gao, and Lihong~V. Wang.
\newblock Single-shot real-time video recording of a photonic mach cone induced
  by a scattered light pulse.
\newblock {\em Science Advances}, 3(1), 2017.

\bibitem{Yao21}
Yunhua Yao, Yilin He, Dalong Qi, Fengyan Cao, Jiali Yao, Pengpeng Ding,
  Chengzhi Jin, Xianyu Wu, Lianzhong Deng, Tianqing Jia, Feng Huang, Jinyang
  Liang, Zhenrong Sun, and Shian Zhang.
\newblock Single-shot real-time ultrafast imaging of femtosecond laser
  fabrication.
\newblock {\em ACS Photonics}, 8(3):738--744, 2021.

\bibitem{Liang18LSA}
Jinyang Liang, Liren Zhu, and Lihong~V. Wang.
\newblock Single-shot real-time femtosecond imaging of temporal focusing.
\newblock {\em Light Sci. Appl}, 7:42, 2018.

\bibitem{Ehn17}
A.~Ehn, J.~Bood, Z.~Li, E.~Berrocal, M.~Aldén, and E.~Kristensson.
\newblock {FRAME}: femtosecond videography for atomic and molecular dynamics.
\newblock {\em Light Sci. Appl}, 6:e17045, 2017.

\bibitem{Qi2020}
Dalong Qi, Shian Zhang, Chengshuai Yang, Yilin He, Fengyan Cao, Jiali Yao,
  Pengpeng Ding, Liang Gao, Tianqing Jia, Jinyang Liang, Zhenrong Sun, and
  Lihong~V. Wang.
\newblock {Single-shot compressed ultrafast photography: a review}.
\newblock {\em Advanced Photonics}, 2(1):1 -- 16, 2020.

\bibitem{Gao14}
L.~Gao, J.~Y. Liang, C.~Y. Li, and L.~V. Wang.
\newblock Single-shot compressed ultrafast photography at one hundred billion
  frames per second.
\newblock {\em Nature}, 516:74--77, 2014.

\bibitem{Lu19}
Yu~Lu, Terence T.~W. Wong, Feng Chen, and Lidai Wang.
\newblock Compressed ultrafast spectral-temporal photography.
\newblock {\em Phys. Rev. Lett.}, 122:193904, May 2019.

\bibitem{Liang20SPCUP}
Jinyang Liang, Peng Wang, Liren Zhu, and Lihong~V. Wang.
\newblock Single-shot stereo-polarimetric compressed ultrafast photography for
  light-speed observation of high-dimensional optical transients with
  picosecond resolution.
\newblock {\em Nature Commun.}, 11:5252, 2020.

\bibitem{Yang20}
Chengshuai Yang, Fengyan Cao, Dalong Qi, Yilin He, Pengpeng Ding, Jiali Yao,
  Tianqing Jia, Zhenrong Sun, and Shian Zhang.
\newblock Hyperspectrally compressed ultrafast photography.
\newblock {\em Phys. Rev. Lett.}, 124:023902, Jan 2020.

\bibitem{Kim20}
Taewoo Kim, Jinyang Liang, Liren Zhu, and Lihong~V. Wang.
\newblock Picosecond-resolution phase-sensitive imaging of transparent objects
  in a single shot.
\newblock {\em Science Advances}, 6(3), 2020.

\bibitem{Wang20}
Peng Wang, Jinyang Liang, and Lihong~V. Wang.
\newblock Single-shot ultrafast imaging attaining 70 trillion frames per
  second.
\newblock {\em Nature Commun.}, 11:2091, 2020.

\bibitem{Qi2021}
Dalong Qi, Fengyan Cao, Shuwu Xu, Yunhua Yao, Yilin He, Jiali Yao, Pengpeng
  Ding, Chengzhi Jin, Lianzhong Deng, Tianqing Jia, Jinyang Liang, Zhenrong
  Sun, and Shian Zhang.
\newblock 100-trillion-frame-per-second single-shot compressed ultrafast
  photography via molecular alignment.
\newblock {\em Phys. Rev. Applied}, 15:024051, Feb 2021.

\bibitem{Kakue12}
T.~{Kakue}, K.~{Tosa}, J.~{Yuasa}, T.~{Tahara}, Y.~{Awatsuji}, K.~{Nishio},
  S.~{Ura}, and T.~{Kubota}.
\newblock Digital light-in-flight recording by holography by use of a
  femtosecond pulsed laser.
\newblock {\em IEEE Journal of Selected Topics in Quantum Electronics},
  18(1):479--485, 2012.

\bibitem{Li14}
Zhengyan Li, Rafal Zgadzaj, Xiaoming Wang, Yen-Yu Chang, and Michael~C. Downer.
\newblock Single-shot tomographic movies of evolving light-velocity objects.
\newblock {\em Nature Commun.}, 5:3085, 2014.

\bibitem{Yue17}
Qing-Yang Yue, Zhen-Jia Cheng, Lu~Han, Yang Yang, and Cheng-Shan Guo.
\newblock One-shot time-resolved holographic polarization microscopy for
  imaging laser-induced ultrafast phenomena.
\newblock {\em Opt. Express}, 25(13):14182--14191, Jun 2017.

\bibitem{Nakagawa14}
K.~Nakagawa, A.~Iwasaki, Y.~Oishi, R.~Horisaki, A.~Tsukamoto, A.~Nakamura,
  K.~Hirosawa, H.~Liao, T.~Ushida, K.~Goda, F.~Kannari, and I.~Sakuma.
\newblock Sequentially timed all-optical mapping photography ({STAMP}).
\newblock {\em Nature Photon.}, 8:695--700, 2014.

\bibitem{Suzuki15}
Takakazu Suzuki, Fumihiro Isa, Leo Fujii, Kenichi Hirosawa, Keiichi Nakagawa,
  Keisuke Goda, Ichiro Sakuma, and Fumihiko Kannari.
\newblock Sequentially timed all-optical mapping photography (stamp) utilizing
  spectral filtering.
\newblock {\em Opt. Express}, 23(23):30512--30522, Nov 2015.

\bibitem{Suzuki17}
Takakazu Suzuki, Ryohei Hida, Yuki Yamaguchi, Keiichi Nakagawa, Toshiharu
  Saiki, and Fumihiko Kannari.
\newblock Single-shot 25-frame burst imaging of ultrafast phase transition of
  ge2sb2te5 with a sub-picosecond resolution.
\newblock {\em Applied Physics Express}, 10(9):092502, aug 2017.

\bibitem{Tournois97}
Pierre Tournois.
\newblock Acousto-optic programmable dispersive filter for adaptive
  compensation of group delay time dispersion in laser systems.
\newblock {\em Optics Communications}, 140(4):245--249, 1997.

\bibitem{Canova09}
Lorenzo Canova, Xiaowei Chen, Alexandre Trisorio, Aur\'{e}lie Jullien, Andreas
  Assion, Gabriel Tempea, Nicolas Forget, Thomas Oksenhendler, and Rodrigo
  Lopez-Martens.
\newblock Carrier-envelope phase stabilization and control using a transmission
  grating compressor and an {AOPDF}.
\newblock {\em Opt. Lett.}, 34(9):1333--1335, May 2009.

\bibitem{Forget09}
N.~Forget, L.~Canova, X.~Chen, A.~Jullien, and R.~Lopez-Martens.
\newblock Closed-loop carrier-envelope phase stabilization with an
  acousto-optic programmable dispersive filter.
\newblock {\em Opt. Lett.}, 34(23):3647--3649, Dec 2009.

\bibitem{Maksimenka10}
Raman Maksimenka, Patrick Nuernberger, Kevin~F. Lee, Adeline Bonvalet, Jadwiga
  Milkiewicz, Cestmir Barta, Milo\v{s} Klima, Thomas Oksenhendler, Pierre
  Tournois, Daniel Kaplan, and Manuel Joffre.
\newblock Direct mid-infrared femtosecond pulse shaping with a calomel
  acousto-optic programmable dispersive filter.
\newblock {\em Opt. Lett.}, 35(21):3565--3567, Nov 2010.

\bibitem{Seiler15}
Hélène Seiler, Brenna Walsh, Samuel Palato, Alexandre Thai, Vincent
  Crozatier, Nicolas Forget, and Patanjali Kambhampati.
\newblock Kilohertz generation of high contrast polarization states for visible
  femtosecond pulses via phase-locked acousto-optic pulse shapers.
\newblock {\em Journal of Applied Physics}, 118(10):103110, 2015.

\bibitem{delaPaz19}
J.~A. de~la Paz, A.~Bonvalet, and M.~Joffre.
\newblock Frequency-domain two-dimensional infrared spectroscopy using an
  acousto-optic programmable dispersive filter.
\newblock {\em Opt. Express}, 27(4):4140--4146, Feb 2019.

\bibitem{Urbanek16}
B.~Urbanek, M.~M\"oller, M.~Eisele, S.~Baierl, D.~Kaplan, C.~Lange, , and
  R.~Huber.
\newblock Femtosecond terahertz time-domain spectroscopy at 36 k{H}z scan rate
  using an acousto-optic delay.
\newblock {\em Appl. Phys. Lett.}, 108:121101, 2016.

\bibitem{Audier17}
Xavier Audier, Naveen Balla, and Herv\'{e} Rigneault.
\newblock Pump-probe micro-spectroscopy by means of an ultra-fast
  acousto-optics delay line.
\newblock {\em Opt. Lett.}, 42(2):294--297, Jan 2017.

\bibitem{Alshaykh17}
Mohammed~S. Alshaykh, Chien-Sheng Liao, Oscar~E. Sandoval, Gregory Gitzinger,
  Nicolas Forget, Daniel~E. Leaird, Ji-Xin Cheng, and Andrew~M. Weiner.
\newblock High-speed stimulated hyperspectral {R}aman imaging using rapid
  acousto-optic delay lines.
\newblock {\em Opt. Lett.}, 42(8):1548--1551, Apr 2017.

\bibitem{Audier20}
Xavier Audier, Nicolas Forget, and Herv\'{e} Rigneault.
\newblock High-speed chemical imaging of dynamic and histological samples with
  stimulated {R}aman micro-spectroscopy.
\newblock {\em Opt. Express}, 28(10):15505--15514, May 2020.

\bibitem{Idlahcen09}
Sa\"{i}d Idlahcen, Lo\"{i}c M\'{e}\`{e}s, Claude Roz\'{e}, Thierry Girasole,
  and Jean-Bernard Blaisot.
\newblock Time gate, optical layout, and wavelength effects on ballistic
  imaging.
\newblock {\em J. Opt. Soc. Am. A}, 26(9):1995--2004, Sep 2009.

\bibitem{Suzuki20}
Takakazu Suzuki, Hirofumi Nemoto, Kazuki Takasawa, and Fumihiko Kannari.
\newblock 1000-fps consecutive ultrafast 2d-burst imaging with a sub-nanosecond
  temporal resolution by a frequency-time encoding of sf-stamp.
\newblock {\em Appl. Phys. A}, 126:135, 2020.

\bibitem{POKRZYWKA2012}
B.~Pokrzywka, A.~Mendys, K.~Dzierżęga, M.~Grabiec, and S.~Pellerin.
\newblock Laser light scattering in a laser-induced argon plasma:
  Investigations of the shock wave.
\newblock {\em Spectrochimica Acta Part B: Atomic Spectroscopy}, 74-75:24--30,
  2012.
\newblock 6th Euro-Mediterranean Symposium on Laser Induced Breakdown
  Spectroscopy (EMSLIBS 2011).

\bibitem{Shaw:06}
Weidong Yang, Alexander~B. Kostinski, and Raymond~A. Shaw.
\newblock Phase signature for particle detection with digital in-line
  holography.
\newblock {\em Opt. Lett.}, 31(10):1399--1401, May 2006.

\bibitem{picart2015}
Pascal Picart.
\newblock {\em New techniques in digital holography}.
\newblock John Wiley \& Sons, 2015.

\bibitem{mazumdar2020}
Yi~Chen Mazumdar, Michael~E Smyser, Jeffery~D Heyborne, Mikhail~N Slipchenko,
  and Daniel~R Guildenbecher.
\newblock Megahertz-rate shock-wave distortion cancellation via phase conjugate
  digital in-line holography.
\newblock {\em Nature communications}, 11(1):1--10, 2020.

\bibitem{Onural93}
Levent Onural.
\newblock Diffraction from a wavelet point of view.
\newblock {\em Opt. Lett.}, 18(11):846--848, Jun 1993.

\bibitem{Buraga2000}
Cristina Buraga-Lefebvre, Sébastien Coëtmellec, Denis Lebrun, and Cafer
  Özkul.
\newblock Application of wavelet transform to hologram analysis:
  three-dimensional location of particles.
\newblock {\em Optics and Lasers in Engineering}, 33(6):409--421, 2000.

\end{thebibliography}

\end{document}